\documentstyle[preprint,revtex]{aps}
\begin{document}
\draft

\begin{title}
Comment on ``Shadow model for sub--barrier fusion applied\\
 to light systems''
\end{title}
\author{C.~A.~Barnes, S.~E.~Koonin, and K.~Langanke}
\begin{instit}
W.~K.~Kellogg Radiation Laboratory, 106--38, Pasadena, CA 91125, USA
\end{instit}

\begin{abstract}
We demonstrate that the cross sections derived from the ``shadow
model'' for reactions between light nuclei disagree with low--energy
laboratory data and exhibit unphysical behavior at energies below
those for which data exist.  As a consequence, the large
thermonuclear reaction rates obtained by Scalia and Figuera [Phys.
Rev. {\bf C46}, 2610 (1992)] are wrong.
\end{abstract}


\narrowtext

In a recent publication \cite{Scalia}, Scalia and Figuera argue that
the rates of the nuclear reactions important in solar hydrogen
burning are substantially larger than those adopted in the Standard
Solar Model \cite{Bahcall}.   This claim is based on a ``shadow
model'' for the energy dependence of the low--energy cross sections.
We demonstrate in this comment that this energy dependence is both
incorrect and unphysical.

In many astrophysical scenarios (e.g., our sun), charged--particle
nuclear reactions proceed at such low energies that a direct
experimental determination of the cross section is not possible with
existing techniques.  Extrapolation of the measured cross sections to
stellar energies is thus necessary.  To be trustworthy, such
extrapolations should not only be tied closely to experimental
information, but should also be guided by a strong theoretical
foundation.

For non--resonant reactions of charged particles (e.g., those that
take place in solar hydrogen burning), tunneling through the Coulomb
barrier dominates the energy dependence of the cross section at the
low energies of astrophysical interest, giving rise to a very rapid
decrease of the cross section $\sigma(E)$ with decreasing
center--of--mass energy $E$.  For a reliable extrapolation,  this
dominant energy dependence is factored out and the cross section is
usually expressed in terms of the astrophysical $S$--factor:
\begin{equation}
S(E)\equiv\sigma(E)\cdot E\cdot \exp\{2\pi\eta(E)\}\;.
\end{equation}
The Sommerfeld parameter is given by
\begin{equation}
\eta(E)={Z_1Z_2e^2\over\hbar v}\;,
\end{equation}
where $v$ is the relative velocity in the entrance channel and
$Z_1,Z_2$ are the charge numbers of the colliding nuclei.  The form
of Eq.~(1) embodies the $s$--wave tunneling through the Coulomb
barrier of two point--like nuclei.  In the absence of near--threshold
resonances, the energy dependence of the $S$--factor is expected to
be weak, reflecting only effects like the strong interaction between
the collision partners, their finite sizes, contributions from other
partial waves, the final state phase space, etc.

The physical picture behind the definition (1) has been confirmed in
numerous measurements of cross sections for reactions between the
light nuclei \cite{Fowler}.  As a typical example, Fig.~1 shows the
astrophysical $S$--factor for the ${}^3{\rm He}({}^3{\rm
He},2p)^4{\rm He}$ reaction, which terminates the $pp$I--chain in
solar hydrogen burning.  The $S$--factor data \cite{Krauss} clearly
show only a very weak and smooth energy dependence indicating that
the $s$--wave penetrability through the Coulomb barrier correctly
describes the low--energy cross section.

These empirical observations are confirmed in a microscopic study
\cite{Typel} of the low--energy ${}^3{\rm He}({}^3{\rm He},2p)^4{\rm
He}$ reaction in which the effects of nuclear structure, the strong
interaction, antisymmetrization, etc. were taken into account.  As
indicated by the solid curve, the parameter--free calculated
energy--dependence of the $S$--factor accurately describes the data.
Thus, one has some confidence that this more elaborate nuclear model
is also capable of extrapolating the astrophysical $S$--factor to the
most effective energy under solar conditions ($E_0\approx 22$~keV).
The calculation yields $S(0)\approx 5.3~\hbox{MeV $\cdot$  b}$, in
close agreement with the value used in the standard solar model
\cite{Bahcall}.  This same microscopic model simultaneously (without
parameter adjustment) reproduces the measured $S$--factors of the
analogue ${}^3{\rm H}({}^3{\rm H},2n)^4{\rm He}$ reaction,
demonstrating that the conventional Gamow barrier penetration
accounts correctly for the physics of low--energy nuclear reactions.

Applying their shadow model for sub--barrier fusion, Scalia and
Figuera \cite{Scalia} obtained low--energy cross sections (and
consequently reaction rates at solar temperatures) that are
significantly higher than the standard values \cite{Bahcall}.
Rather than being based on the correct physical picture of barrier
penetration, the energy dependence of the low--energy cross section
in this model is simply assumed (Eqs. (3--5) and (8,9) in
Ref.~\cite{Scalia}).

We will demonstrate that this assumption is wrong.  Having six fit
parameters at their disposal, Scalia and Figuera claim \cite{Scalia}
to reproduce the energy dependence of the measured cross sections,
and support their claim by numerous figures in which cross sections
are plotted as functions of $E$.  However, as these figures use a
logarithmic scale to plot the rapidly--varying cross sections, it is
difficult to judge the success of the shadow model approach  in
reproducing the data.  To do so more easily, we have transformed the
fusion cross sections as calculated from the shadow model Eqs. (3--9)
of Ref.~\cite{Scalia} (parameters as given in Table~I of that
reference) into the $S$-factor defined by (1).  As a typical example,
we compare the ${}^3{\rm He}({}^3{\rm He},2p)^4{\rm He}$ $S$--factors
predicted by Scalia and Figuera with the most modern and precise
data \cite{Krauss}.  It is obvious from our figure that the shadow
model does not reproduce the energy dependence of the measured
${}^3{\rm He}({}^3{\rm He},2p)^4{\rm He}$ data.  More importantly,
there is an unjustified and unphysical increase of the $S$--factor at
very low energies, leading to the large shadow model reaction rates.

We find similar inaccuracies for the other reactions considered in
Ref.~\cite{Scalia} and, in each case, the model predicts an
unphysical, dramatic increase of the $S$--factor at energies smaller
than those for which data are available.  We therefore conclude that
the shadow model is not useful for extrapolating measured cross
sections to astrophysically relevant energies, and that any
conclusions drawn from such extrapolations are unjustified.

\acknowledgments
This work was supported in part by the National Science Foundation,
Grant Nos. PHY90--13248 and PHY91--15574.

\figure{$S$--factor for the ${}^3{\rm He}({}^3{\rm He},2p){}^4{\rm
He}$ reaction.  Points are the experimental data from
Ref.~\cite{Krauss}.  The solid line is the energy dependence
predicted by the microscopic model of Ref.~\cite{Typel}, while the
dashed curve shows the shadow model prediction of Ref.~\cite{Scalia}.
The arrow indicates $E_0$, the ``most effective energy'' for a
temperature of $15\times 10^6$~K.}

\end{document}